\def\gsim{~\rlap{$>$}{\lower 1.0ex\hbox{$\sim$}}}
\def\lsim{~\rlap{$<$}{\lower 1.0ex\hbox{$\sim$}}}

\def\wpm2{W m$^{-2}$}

\def\eg{{\it e.g.\ }}

\newcounter{foo}
\documentclass{emulateapj}

\usepackage{CJKutf8}

\begin{document}
\begin{CJK}{UTF8}{bsmi}


\title{VERY LOW-MASS STELLAR AND SUBSTELLAR COMPANIONS TO SOLAR-LIKE STARS FROM MARVELS III: A SHORT-PERIOD BROWN DWARF CANDIDATE AROUND AN ACTIVE G0IV SUBGIANT}

\author{Bo Ma(馬波)\altaffilmark{1}, 
Jian Ge\altaffilmark{1}, 
Rory Barnes\altaffilmark{2}, 
Justin R. Crepp\altaffilmark{3,4},
Nathan De Lee\altaffilmark{1,8},
Leticia Dutra-Ferreira \altaffilmark{11,24},
Massimiliano Esposito\altaffilmark{5,6},
Bruno Femenia\altaffilmark{5,6},
Scott W. Fleming\altaffilmark{9,22,1}, 
B. Scott Gaudi\altaffilmark{7},
Luan Ghezzi\altaffilmark{10,24},
Leslie Hebb\altaffilmark{8},
Jonay I. Gonzalez Hernandez\altaffilmark{5,6},
Brian L. Lee\altaffilmark{1,2},
G. F. Porto de Mello\altaffilmark{11,24},
Keivan G. Stassun\altaffilmark{8,19}, 
Ji Wang\altaffilmark{1},
John P. Wisniewski\altaffilmark{26},
Eric Agol\altaffilmark{2},
Dmitry Bizyaev\altaffilmark{14},
Phillip Cargile\altaffilmark{8},
Liang Chang\altaffilmark{1},
Luiz Nicolaci da Costa\altaffilmark{10,24},
Jason D. Eastman\altaffilmark{7,20,21}, 
Bruce Gary\altaffilmark{8},
Peng Jiang\altaffilmark{1},
Stephen R. Kane\altaffilmark{13},
Rui Li\altaffilmark{1},
Jian Liu\altaffilmark{1},
Suvrath Mahadevan\altaffilmark{1,9,22},
Marcio A. G. Maia\altaffilmark{10,24},
Demitri Muna\altaffilmark{18},
Duy Cuong Nguyen\altaffilmark{1},
Ricardo L. C. Ogando\altaffilmark{10,24},
Daniel Oravetz\altaffilmark{14},
Joshua Pepper\altaffilmark{8},
Martin Paegert\altaffilmark{8},
Carlos Allende Prieto\altaffilmark{5,6},
Rafael Rebolo\altaffilmark{5,23},
Basilio X. Santiago\altaffilmark{24,25},
Donald P. Schneider\altaffilmark{9,22},
Alaina Shelden\altaffilmark{14},
Audrey Simmons\altaffilmark{14},
Thirupathi Sivarani\altaffilmark{1,12},
J. C. van Eyken\altaffilmark{13,21},
Xiaoke Wan\altaffilmark{1},
Benjamin A. Weaver\altaffilmark{18},
Bo Zhao\altaffilmark{1},
}
\email{boma@astro.ufl.edu}
\altaffiltext{1}{Department of Astronomy, University of Florida, 211 Bryant Space Science Center, Gainesville, FL, 32611-2055, USA}
\altaffiltext{2}{Department of Astronomy, University of Washington, Box 351580, Seattle, WA 98195-1580, USA}
\altaffiltext{3}{Department of Physics, University of Notre Dame, 225 Nieuwland Science Hall, Notre Dame, IN 46556, USA}
\altaffiltext{4}{Department of Astronomy, California Institute of Technology, 1200 E. California Blvd, Pasadena, CA 91125, USA}
\altaffiltext{5}{Instituto de Astrof\'{\i}sica de Canarias, C/V\'{\i}a Láctea S/N, E-38200 La Laguna, Spain}
\altaffiltext{6}{Departamento de Astrof\'{\i}sica, Universidad de La Laguna, E-38205 La Laguna, Tenerife, Spain}
\altaffiltext{7}{Department of Astronomy, The Ohio State University, 140 West 18th Avenue, Columbus, OH 43210, USA}
\altaffiltext{8}{Department of Physics and Astronomy, Vanderbilt University, Nashville, TN 37235, USA}
\altaffiltext{9}{Department of Astronomy and Astrophysics, The Pennsylvania State University, 525 Davey Laboratory, University Park, PA 16802, USA}
\altaffiltext{10}{Observat\'{o}rio Nacional, Rua General Jos\'{e} Cristino, 77, 20921-400 S\~{a}o Crist\'{o}v\~{a}o, Rio de Janeiro, RJ, Brazil}
\altaffiltext{11}{Universidade Federal do Rio de Janeiro, Observat\'{o}rio do Valongo, Ladeira do Pedro Ant\^{o}nio, 43, CEP: 20080-090, Rio de Janeiro, RJ, Brazil}
\altaffiltext{12}{Indian Institute of Astrophysics, II Block, Koramangala, Bangalore 560 034, India}
\altaffiltext{13}{NASA Exoplanet Science Institute, Caltech, MS 100-22, 770 South Wilson Avenue, Pasadena, CA 91125, USA}
\altaffiltext{14}{Apache Point Observatory, P.O. Box 59, Sunspot, NM 88349-0059, USA}
\altaffiltext{15}{MIT Kavli Institute for Astrophysics \& Space Research, Cambridge, MA 02139, USA}
\altaffiltext{16}{Steward Observatory, University of Arizona, Tucson, AZ 85121, USA}
\altaffiltext{17}{Department of Astronomy, MSC 4500, New Mexico State University, P.O. Box 30001, Las Cruces, NM 88003, USA}
\altaffiltext{18}{Center for Cosmology and Particle Physics, New York University, New York, NY, USA}
\altaffiltext{19}{Department of Physics, Fisk University, 1000 17th Ave. N., Nashville, TN 37208, USA}
\altaffiltext{20}{Las Cumbres Observatory Global Telescope Network, 6740 Cortona Drive, Suite 102, Santa Barbara, CA 93117, USA}
\altaffiltext{21}{Department of Physics Broida Hall, University of California, Santa Barbara, CA 93106, USA}
\altaffiltext{22}{Center for Exoplanets and Habitable Worlds, The Pennsylvania State University, University Park, PA 16802, USA}
\altaffiltext{23}{Consejo Superior de Investigaciones Científicas, Spain}
\altaffiltext{24}{Laborat\'{o}rio Interinstitucional de e-Astronomia (LIneA), Rio de Janeiro, RJ 20921-400, Brazil}
\altaffiltext{25}{Instituto de F\'{\i}sica, UFRGS, Caixa Postal 15051, Porto Alegre, RS 91501-970, Brazil}
\altaffiltext{26}{Homer L Dodge Department of Physics \& Astronomy, University of Oklahoma, 440 W Brooks St, Norman, OK 73019, USA}

\begin{abstract}
We present an eccentric, short-period brown dwarf 
candidate orbiting the active, slightly evolved subgiant star 
TYC~2087-00255-1, which has effective temperature 
$T_{\rm eff} = 5903 \pm 42$\,K, surface gravity $\log (g) = 4.07 \pm 0.16$\,(cgs), 
and metallicity $\rm [Fe/H] = -0.23 \pm 0.07$. 
This candidate was discovered using data from the first two 
years of the Multi-object APO Radial Velocity Exoplanets Large-area 
Survey (MARVELS), which is part of the third phase of Sloan Digital Sky 
Survey. From our 38 radial velocity measurements spread 
over a two-year time baseline, we derive a Keplerian orbital 
fit with semi-amplitude $K=3.571 \pm 0.041$\,km~s$^{-1}$, period 
$P=9.0090 \pm 0.0004$\,days, and eccentricity $e=0.226\pm 0.011$. 
Adopting a mass of $1.16 \pm 0.11\,M_\odot$ for the subgiant host star, 
we infer that the companion has a minimum mass of $40.0 \pm 2.5\,M_{Jup}$. 
Assuming an edge-on orbit, the semimajor axis is $0.090 \pm 0.003$\,AU. 
The host star is photometrically variable at the $\sim 1\%$ level with a 
period of $\sim 13.16 \pm 0.01$ days, indicating that the host star spin 
and companion orbit are not synchronized. Through adaptive 
optics imaging we also found a point source $643\pm10$ mas away from 
TYC~2087-00255-1, which would have a mass of $0.13M_\odot$ 
if it is physically associated with TYC~2087-00255-1 and has the 
same age. Future proper motion observation should be able 
to resolve if this tertiary object is physically associated with 
TYC~2087-00255-1 and make TYC~2087-00255-1 a triple body 
system. Core  Ca~II H and K line emission indicate that the host is 
chromospherically active, at a level that is consistent with the inferred 
spin period and measured $v_{rot}\sin i$, but unusual for a subgiant of 
this $T_{\rm eff}$. This activity could be explained by ongoing tidal spin-up 
of the host star by the companion.
\end{abstract}

\section{INTRODUCTION} 

Brown dwarfs (BDs) range in mass from $\sim$13$-$80 Jupiter masses and burn deuterium 
but not hydrogen \citep{burrows97, chabrier00, burrows01, spiegel11}. 
The first unambiguous discovery of BDs \citep{rebolo95,nakajima95, oppenheimer95,basri96,rebolo96} 
occurred at the same time as the discovery of the first extra-solar giant planet orbiting a main 
sequence star \citep[51 Peg b;][]{mayor95}. More than 800 BDs have been discovered to date 
(see DwarfArchives.org, http://www.dwarfarchives.org). Most of them are free-floating 
objects and only several dozen BDs are companions to other stars \citep{reid08,sahlmann11}.
The BD desert (a paucity of BD companions relative to planetary or 
stellar companions within 3 AU around main-sequence FGKM stars) 
was found during high-precision radial velocity (RV) surveys seeking 
exoplanets \citep{marcy00}. Since RV surveys are more sensitive to BDs than to exoplanets, this 
paucity is a real minimum in the mass distribution of close companions to solar-type stars. 
The California \& Carnegie Planet Search finds a BD occurrence rate of 
0.7\% $\pm$ 0.2\% from their sample of $\sim \! 1000$ target stars \citep{Vogt02, Patel07}, 
and the McDonald Observatory Planet Search agrees, with a rate of 0.8\% $\pm$ 0.6\% from a search
sample of 250 stars \citep{Wittenmyer09}.  To assess the reality of the BD desert, 
\citet{Grether06} performed a detailed investigation of the companions around nearby 
Sun-like stars. They find that approximately $16\%$ of nearby Sun-like stars have close ($P < 5$~yr) 
companions more massive than Jupiter: 11$\%\pm3\%$ 
are stellar, $<1\%$ are BDs, and $5\%\pm2\%$ are giant planets.
However, \citet{Gizis01} suggest that BDs might not be as rare at wide 
separations \citep[see also][]{Metchev04}. \citet{laf07} obtain a $95\%$ confidence 
interval of $1.9^{+8.3}_{-1.5}\%$ for the frequency of 13$-$75$M_{\rm Jup}$ 
companions between 25$-$250AU amongst 85 nearby young stars observed 
during the Gemini Deep Planet Survey. Based on an adaptive optics survey 
for substellar companions, \citet{metchev09} infer that the frequency 
of BDs in 28$-$1590 AU orbits around young solar analogs is $3.2^{+3.1}_{-2.7}\%$. 

Ostensibly, BDs are believed to form similarly to stars, through gravitational collapse and/or fragmentation of 
molecular clouds \citep{Padoan04, Hennebelle08}. However, companions with 
masses up to 10 $M_{\rm Jup}$ \citep{Alibert05} or even 25 $M_{\rm Jup}$ 
\citep{Mordasini08} may form in protoplanetary disks. As such, the BD 
desert is commonly interpreted as the gap between the largest mass objects that 
can be formed in disks and the smallest mass clump that can 
collapse and/or fragment in the vicinity of a protostar. 
In comparison, the mass function of isolated substellar objects both in the field and in clusters 
appears to be roughly flat in $\log(\rm M)$ for masses down to at least $\sim\!20\,M_{Jup}$ \citep{luhman00,chabrier02}. Recently, \citet{Andre12} found a self-gravitating condensation of gas and dust with a mass of 0.015 to 0.03$M_\odot$ using millimeter interferometric observations, 
which supports the idea that BDs could form the same way as stars.

Given the occurrence rate of $\sim \! 1\%$ for BD companions, a large, relatively uniform,
systematic RV survey of a much larger sample of stars is needed to make 
further progress in understanding properties of the BD desert. The Multi-object APO 
Radial Velocity Exoplanets Large-area Survey \citep[MARVELS;  ][]{Ge08} 
is a four-year RV survey of $\sim \! 3,300$ stars with 
$7.6 \! < \! V \! < \! 12$ over time baselines of $\sim \! 1.5$ years per target, 
with a stated goal of $< \! 30$\,m~s$^{-1}$ precision for the faintest stars. 
MARVELS uses the innovative instrumental technique of a dispersed 
fixed-delay interferometer \citep[DFDI;  see, e.g.,][]{erskine00,Ge02,Ge09,vaneyken10, wang11, 
wang12a,wang12b} 
in order to simultaneously observe 60 objects at a time. By virtue of the large 
number of target stars, as well as the addition of uniform selection criteria 
described in \citet{lee11}, MARVELS is well-suited to detect significant numbers 
of rare companions. For example, \citet{lee11} have recently announced 
MARVELS-1b, a 5.9 day BD candidate around a F-type star 
TYC 1240- 00945-1 located in the BD desert.

This paper is part of a series that describes the very low-mass stellar and substellar 
companions to solar-like stars detected in the MARVELS survey 
\citep{lee11,wis12,fleming12}. In this paper, we report a new MARVELS BD candidate, 
which we designate MARVELS-4b, detected in orbit around the star 
TYC~2087-00255-1 ($Tycho$-2 star catalog;  \citealt{hog00}). In \S 2 we 
describe the observations used in this paper. We present 
stellar parameters for the star in \S 3 and orbital parameters for the BD candidate in 
\S 4. We discuss these results and give our conclusions in \S 5.

\section{OBSERVATIONS AND PROCESSING} 

\subsection{MARVELS Radial Velocities}
\label{sec:sdssobs} 

TYC~2087-00255-1 was a target in the first two-year cycle of the SDSS-III \citep{Eisenstein11} 
MARVELS planet search program. This star was selected for RV 
monitoring using the preselection methodology and instrumentation described in \citet{lee11}.  
The RV observations were taken using the Sloan 
Digital Sky Survey (SDSS) 2.5-m telescope at Apache 
Point Observatory \citep{gunn06} coupled to the MARVELS instrument \citep{Ge09}, 
a 60 object, fiber-fed, dispersed fixed-delay interferometer. The interferometer produces two 
fringing spectra (``beams'') per object, in the wavelength range $\sim \! 500-570$\,nm, 
with resolving power $R \! \sim \! 12000$. TYC~2087-00255-1 was observed at 23 
epochs from 2009 May 4 to 2010 July 5. \citet{lee11} describe the basic data 
reduction and analysis leading to the production of differential RVs. 
The RV errors are scaled by a ``quality factor'' $Q=6.22$ based on the rms 
errors of the other stars observed on the same SDSS-III plate as 
TYC~2087-00255-1 \citep{fleming10}.
The differential RV measurements for TYC~2087-00255-1 from MARVELS 
are summarized in Table~\ref{tab:obsjournalhet1}. Note that a constant velocity term has 
been subtracted to account for the instrument offset (see \S \ref{sec:kpo} for more detail).

\begin{table}[htbp]
\begin{center}
\caption{{SUMMARY of MARVELS RADIAL VELOCITIES \label{tab:obsjournalhet1}}}
\begin{tabular}{lcc}
\hline\hline
HJD  & RV & $\sigma_{RV}$ \\
&  km~s$^{-1}$  & km~s$^{-1}$     \\
\hline
 2454955.86681  & -1.728 &  0.056 \\
 2454995.81378  &  3.007 &  0.048 \\
 2455020.70639  & -3.583 &  0.045 \\
 2455025.65578  &  2.354 &  0.053 \\
 2455106.66179  &  2.333 &  0.053 \\
 2455107.63962  &  1.155 &  0.053 \\
 2455255.00502  & -3.105 &  0.094 \\
 2455259.00872  &  3.111 &  0.045 \\
 2455281.89957  & -3.211 &  0.063 \\
 2455284.87791  &  3.465 &  0.059 \\
 2455286.89844  &  2.444 &  0.051 \\
 2455291.86506  & -0.911 &  0.053 \\
 2455292.87266  &  2.390 &  0.057 \\
 2455311.89105  &  3.574 &  0.053 \\
 2455312.80083  &  3.190 &  0.061 \\
 2455314.80872  &  1.226 &  0.055 \\
 2455338.78513  &  3.564 &  0.044 \\
 2455346.71409  &  1.869 &  0.095 \\
 2455347.70953  &  3.507 &  0.072 \\
 2455350.83978  &  0.905 &  0.067 \\
 2455374.88651  &  3.536 &  0.052 \\
 2455381.88436  & -1.373 &  0.061 \\
 2455382.91902  &  2.255 &  0.065 \\
\hline
\end{tabular}
\end{center}
\end{table}

\subsection{Spectra For Stellar Characterization}


In pursuit of precise stellar parameters for the primary, optical 
($\sim \! 3500-9000$\,\AA) spectra of TYC~2087-00255-1 were obtained using the 
FEROS high resolution ($R \! = \! 48000$) spectrograph \citep{Kaufer99} mounted 
on the MPG/ESO 2.2-m telescope in La Silla on 2010 August 2. FEROS spectra were 
analyzed using the online FEROS Data Reduction System (DRS) and 
the standard calibration plan, where bias, flat-field and wavelength 
calibration lamp frames are observed in the afternoon. 
Three 2400s exposures were combined to yield a S/N of 
$\sim230$ per one-dimensional extracted pixel at $6600$\,\AA.

Additional spectroscopic observations around the H$\alpha$ line were also 
secured with the coud\'e spectrograph of the 1.60m telescope at
Observat\'orio do Pico dos Dias, Brazil on 2010 August 17. 
The resolution was set to R =18\,000 and the S/N per 
pixel was $\sim$ 80. Data reduction was carried out by 
the standard procedure using IRAF. After usual bias and
flat-field correction, we subtracted the background and scattered
light and extracted one-dimensional spectra. No fringing was present
in our spectra. The OPD coud\'e single order spectrograph introduces
no necessity for blaze function corrections and thus the line profile
is easily normalized, lending itself to accurate analysis of the
temperature profile of the stellar atmosphere by fitting the observed
profile to theoretical calculations. This will be used below to infer
an independent $T_{\rm eff}$ estimate as well as the evaluation of the
chromospheric radiative losses in the H$\alpha$ line core.


\subsection{Additional Radial Velocity Observations}
\label{sec:rvfollowup}

High-resolution spectra were collected with the Spectrografo di Alta 
Resoluzione Galileo (SARG) spectrograph \citep{gratton01} 
at the 3.58m Telescopio Nazionale Galileo (TNG) for additional RV 
measurements from 2010 August 27 to 2011 August 17. 
This spectrograph provides $R \sim 57,000$ spectra 
spanning a wavelength range of 462$-$792 nm. The spectra were reduced 
using standard IRAF Echelle reduction packages. Frames were trimmed, 
bias subtracted, flat-field corrected, aperture-traced and extracted. We 
obtained 15 epochs of observations with an iodine cell and an 
additional epoch without the iodine cell to serve as a stellar template. The 
exposure time for each epoch is 20 to 30 minutes. 
The signal-to-noise ratio (S/N) per resolution element at 550 nm is $\sim 60-250$, 
and one resolution element is sampled by 4.9 pixels.
A total of 15 RV data points were derived using the iodine cell technique 
\citep{marcy00}. Each of 21 SARG orders between 504 and 611 nm 
was subdivided into 10 sections, and then RVs were measured from these 
components. Following a $2-\sigma$ clip, the measurements were averaged to 
produce the RVs. The resultant differential RVs are summarized in Table~2. 
Note that a constant velocity term has been subtracted to account for the instrument 
offset (see \S \ref{sec:kpo})

\begin{table}[htbp]
\begin{center}
\caption{{SUMMARY of SARG RADIAL VELOCITIES \label{tab:obsjournalhet2}}}
\begin{tabular}{lcc}
\hline\hline
HJD  & RV & $\sigma_{RV}$ \\
 & km~s$^{-1}$  & km~s$^{-1}$     \\
\hline
 2455436.40537  &  0.679 &  0.046 \\
 2455436.49082  &  0.960 &  0.066 \\
 2455460.36043  & -1.475 &  0.019 \\
 2455460.38315 & -1.493 &  0.018 \\
 2455495.32045  &  0.467 &  0.068 \\
 2455495.33502  &  0.318 &  0.096 \\
 2455495.35302 &  0.259 &  0.137 \\
 2455666.65770 &    0.096 &  0.044 \\
 2455698.58443 &     3.028 &  0.010 \\
 2455725.45470 &   2.654 &  0.022 \\
 2455725.48125 &    2.701 &  0.020 \\
 2455725.53485 &     2.801 &  0.021 \\
 2455760.43524 & -0.550 &  0.035 \\
 2455760.56766 &    0.133 &  0.044 \\
 2455791.50436 &    2.219 &  0.032 \\
\hline
\end{tabular}
\end{center}
\end{table}

%

\subsection{Diffraction-limited Imaging}
High angular resolution lucky images (LIs, observations taken at very high 
cadence to achieve nearly-diffraction-limited 
images from a subsample of the total) were obtained using FastCam \citep{osc2008} on the 
1.5 m TCS telescope at Observatorio del Teide, Spain. The primary goal of these 
observations was to search for companions at large separations that could pollute 
spectroscopic RV observations of the targets. The LI frames were acquired on 
2011 May 8 and 2011 July 1 in the $I$ band, covering $\sim 21\arcsec \times 21\arcsec$ 
on the sky. A total of 60000 images, each corresponding to 60 ms integrations, 
were taken on 2011 May 8, and a total of 64000 frames, each corresponding to 
50 ms integrations, were taken on 2011 July 1. 

To further assess the multiplicity of TYC~2087-00255-1, we acquired AO images using NIRC2 (instrument built by Keith Matthews) on the Keck II telescope on 2012 August 25 UT. 
TYC~2087-00255-1 is bright (V=10.6) and served as its own on-axis natural guide star.
NIRC2 is a high-resolution near-infrared camera that provides a plate scale (when operating in narrow mode) of $9.963\pm0.006$ mas/pix \citep{ghez08} and $10.2^{''}\times10.2^{''}$ field of view. 
Our observations consist of 
dithered frames taken with the K’ band ($\lambda_c=2.12 \mu$m, $\Delta \lambda=0.35 \mu$m) 
filter. 
The total on-source integration time was 47.5 seconds.


\subsection{SuperWASP Photometric Data}
To check for intrinsic photometric variability indicating stellar activity and search 
for possible transits of the companion, we extracted photometric time series data of TYC~2087-00255-1 from the SuperWASP public archive \citep{butters10}. 
The WASP instruments provide flux measurements for millions of stars 
using wide-angle images of the night sky over a band 
pass of 400-700 nm defined by a broad-band filter. Eight cameras on each instrument 
provide images covering approximately $7.8\times7.8$ degrees using Canon 200mm f/1.8 
camera lenses and e2v $2048\times2048$ CCDs. Synthetic aperture photometry using 
an aperture radius of 49 arcseconds at the position of targeted stars is performed on 
the images \citep{Pollacco06}. A total of 18935 aperture photometry data points, 
each taken with a 30 second integration time, were available from the SuperWASP public archive. 
These data were taken between 2004 May 2 and 2008 August 10. Systematic errors  
caused by spatially localized flat-fielding, errors in the vignetting correction near the edge 
of the field of view, bright moonlight contamination, bad weather and other as-yet 
unidentified reasons do exist in SuperWASP data sets \citep{cameron06}. 
There are large systematic errors in some of the data sets, and we choose to 
use data points with relative flux errors smaller than 0.01 in our further analysis, 
to limit systematics. The final selected data sets have a total of 11932 photometric 
data points. 

\section{TYC~2087-00255-1: THE STAR} 
\label{sec:star}
\subsection{Stellar Parameters}
 \label{sec:params} 

\begin{figure}
\plotone{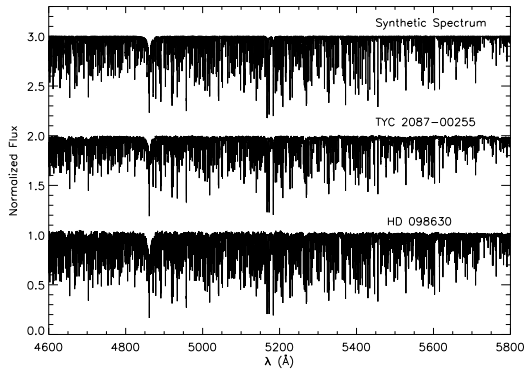}
\caption{Continuum-normalized, high resolution FEROS spectra of TYC~2087-00255-1, 
which is a G0IV star. 
For comparison purpose, also shown in this plot are the spectrum of a 
G0IV star HD 098630 from Elodie spectra library \citep{prugniel01} and the 
synthetic spectrum of TYC~2087-00255-1 calculated using SME package 
(please see the text for more information). The FEROS spectrum and the 
synthetic spectrum have been shifted in the y-axis direction for display purpose.  
\label{fig:type} }
\end{figure}

We have used two independent pipelines (referred to as ``BPG'' and ``IAC'') to derive 
the stellar parameters from the high resolution FEROS spectra. Both methods 
involve analysis of the equivalent widths of Fe I and Fe II lines to 
balance the excitation and ionization equilibria of these features. 
These two methods are described in detail in \citet{wis12}. Both pipelines 
produce values of effective temperature $T_{\rm eff}$, surface gravity $\log (g)$ 
and metallicity $\rm [Fe/H] $ mutually consistent. Because both pipeline determinations are 
mutually consistent, we average these two sets weighted by their own 
pipeline errors to determine the final stellar parameters, shown as `combined 
results' in Table~\ref{tab:stellarparamsraw}. For each stellar 
parameter, we add in quadrature a systematic error of
18 K, 0.08 and 0.03 for $T_{\rm eff}$, $\log (g)$ and $\rm [Fe/H]$, respectively, 
in addition to the internal errors inherent from the two pipeline 
results \citep[see][]{wis12}. The results are summarized in Table~\ref{tab:stellarparamsraw}. 
The adopted stellar parameters are $T_{\rm eff} = 5903 \pm 42$\,K, 
$\log (g) = 4.07 \pm 0.16$\,(cgs), and $\rm [Fe/H] = -0.23 \pm 0.07$. 
So TYC~2875-00255-1 is a G0IV type star. In Fig.~\ref{fig:type} we show
the continuum-normalized high resolution FEROS spectra of TYC~2087-00255-1 together 
with the spectra of HD~098630, a known G0IV type star from Elodie spectra library \citep{prugniel01}. For comparison purpose, we also show synthetic spectra of 
TYC~2087-00255-1 in Fig.~\ref{fig:type}. The synthetic spectra is calculated 
using software package Spectroscopy Made Easy \citep[SME;][]{valenti96} 
and the stellar parameters from the above analysis.

\begin{table*}[htbp]
\begin{center}
\caption{{ Spectroscopic Parameters of the Star TYC~2087-00255-1\label{tab:stellarparamsraw}}}
\begin{tabular}{cccccl}
\hline\hline
$T_{\rm eff}$ & $\log(g)$ & [Fe/H] & $\xi_{t}$ & Notes \\
(K) & (cgs) &  & (km~s$^{-1}$) &  \\
\hline
5805 $\pm$ 71 & 4.02 $\pm$ 0.18 & -0.24 $\pm$ 0.10 & 1.74 $\pm$ 0.10  & BPG (ESO 2.2-m)\\
5941 $\pm$ 44 & 4.15 $\pm$ 0.22 & -0.23 $\pm$ 0.08 & 1.687 $\pm$ 0.055 & IAC (ESO 2.2-m)\\
5903 $\pm$ 42 & 4.07 $\pm$ 0.16 & -0.23 $\pm$ 0.07 & 1.70$\pm$ 0.06 & combined results\\
\hline
\end{tabular}
\end{center}
\end{table*}

\begin{figure}
\plotone{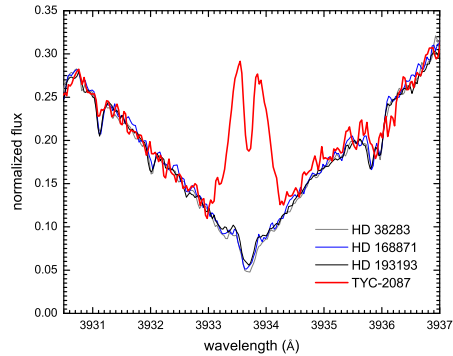}
\caption{Continuum-normalized, high resolution FEROS spectra of TYC~2087-00255-1 
centered on the Ca~II K line. The excess emission from the Ca~II K 
line core indicates that TYC~2087-00255-1 is a chromospherically active star. 
Also shown on this plot are several chromospherically quite stars 
with similar stellar parameters to TYC~2087-00255-1 for comparison purpose 
\citep{ghezzi10a, ghezzi10b}.  \label{fig:HK} }
\end{figure}

Continuum-normalized, high-resolution FEROS spectra of TYC~2087-00255-1 
centered on the Ca~II K line is shown in Fig.~\ref{fig:HK}. Also shown 
in this plot are three stars with similar stellar parameters as TYC~2087-00255-1 
\citep{ghezzi10a, ghezzi10b}. The excess emission from the Ca~II K 
line core indicates that TYC~2087-00255-1 is more active than stars 
with similar stellar parameters. Using the high-resolution 
FEROS spectra and the method described in \citet{jenkins08}, 
the chromospheric Ca~II HK activity index of TYC~2087-00255-1 
is $\log R^{'}_{\rm HK}=-4.58$, with a calibration rms error of 0.03. 

\begin{table}[htbp]
\begin{center}
\caption{Derived Parameters of the Star TYC~2087-00255-1 \label{tab:stellarparams}}
\begin{tabular}{lc}
\hline\hline
Parameter & Value \\
\hline
Spectral Type & G0 IV \\ 
Mass & 1.16 $\pm$ 0.11 $M_{\odot}$\\ 
Radius & $1.64\pm0.37 R_{\odot}$\\ 
Age & 5.5~Gyr \\
$A_V$ & $0.12^{+0.03}_{-0.06}$\\
Distance & $218\pm 14$ pc\\
$\log R^{'}_{\rm HK}$ & -4.58 \\ 
\hline
\end{tabular}
\end{center}
\end{table}

\begin{figure}
\plotone{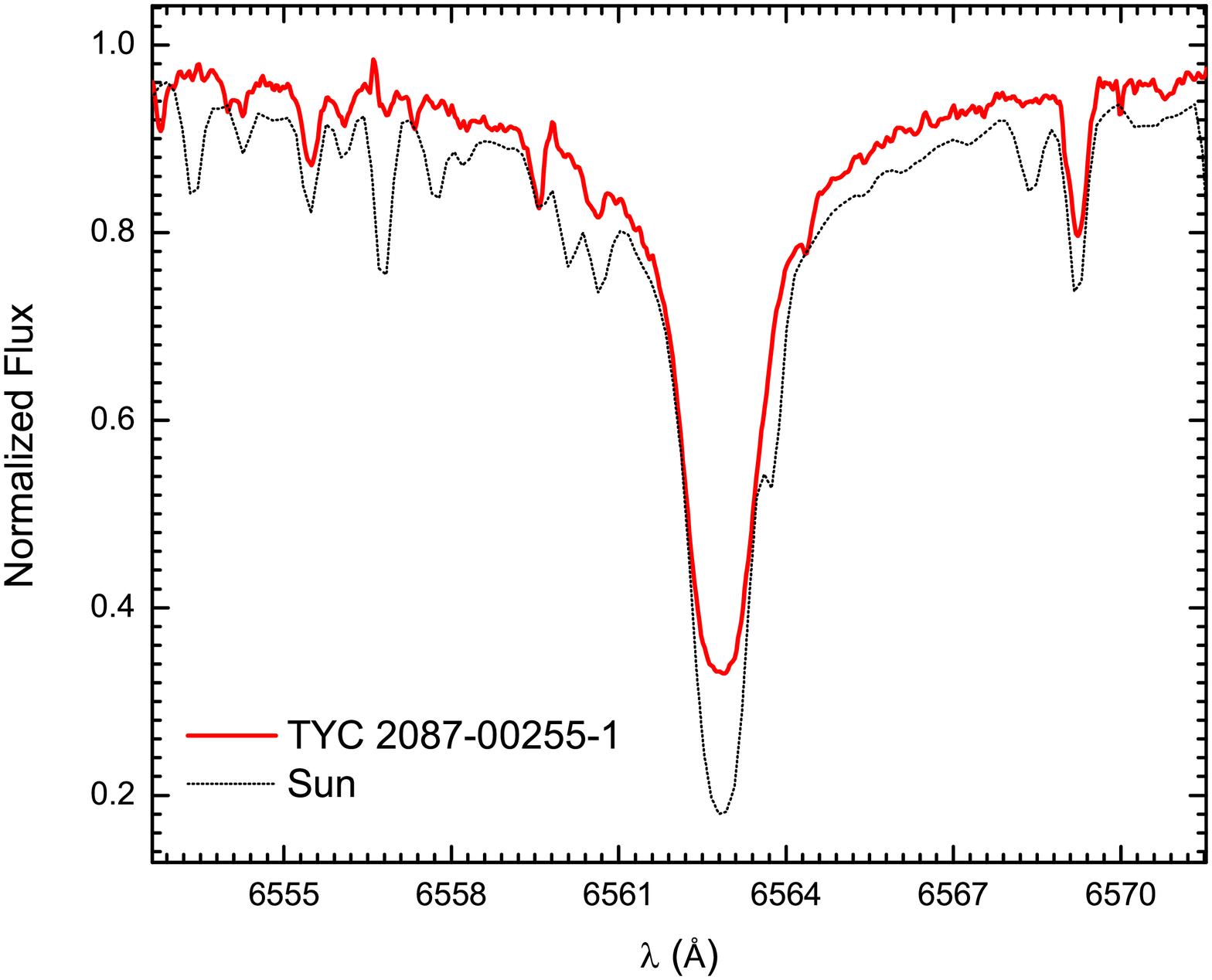}
\caption{\label{fig:halpha} Continuum-normalized H$\alpha$ profile of 
TYC~2087-00255-1, with a spectroscopic resolution of $18\,000$. 
Also shown on the plot is H$\alpha$ profile of the Sun for comparison. 
The solar spectrum is measured from the reflected light of the Sun 
by Ganymede, taken together with TYC~2087-00255-1 using 
the same instrument. The shallower wing profile and the stronger core filling in the H$\alpha$ 
line when compared to the Sun suggests an effective temperature 
lower than the spectroscopic temperature, which is also found by the SED analysis.}
\end{figure}

The H$\alpha$ profile of TYC~2087-00255-1 is shown in Fig.~\ref{fig:halpha} 
superimposed over the solar one, the latter obtained as a disk-integrated 
spectrum from Ganymede (which reflects light from the Sun) 
under the same observational conditions. 
The shallower wing profile is apparent, translating into a lower 
$T_{\rm eff}$ for TYC~2087-00255-1, as is the much stronger line 
core filling, interpreted as additional chromospheric fill-in. Note that 
the H$\alpha$ line  core is substantially broader in TYC~2087-00255-1,
interpreted as yet another confirmation of the subgiant status of 
TYC~2087-00255-1 \citep{pasquini91, lyra05}. The chromospheric 
loss in the H$\alpha$ core of TYC~2087-00255-1 
was also evaluated under the prescription of \citet{lyra05},
using as input the spectroscopic atmospheric parameters. We have
derived a total chromospheric flux of $13.4\times10^5$ erg~cm$^{-2}$~s$^{-1}$ 
inside the H$\alpha$ line core, the estimated error in the \citet{lyra05}
procedure being $\sim0.5\times10^5$ erg~cm$^{-2}$~s$^{-1}$. 
This value is probably also contaminated by veiling from the 
companion, and should be taken as an upper limit. 
Nonetheless, it is over four times the expected flux from
typical subgiants in this $T_{\rm eff}$ range \citep[Fig.~3 of ][]{lyra05} 
and lies between the average chromospheric radiative losses of
Pleiades (age $\sim100$ Myr) and Ursa Major group stars ($\sim400$ Myr). 
This flux is in very good agreement with the Ca~II K line profile (which
being in the UV is probably free from the secondary's contamination)
in pointing to a very high activity level in TYC~2087-00255-1, compatible with a
solar-type stars no more than a few hundred million years of 
age \citep{lyra05}. Two independent spectroscopic chromospheric 
indicators therefore confirm TYC~2087-00255-1 as much more active 
than expected from its subgiant status.

\begin{figure}
\includegraphics[angle=90,scale=0.35]{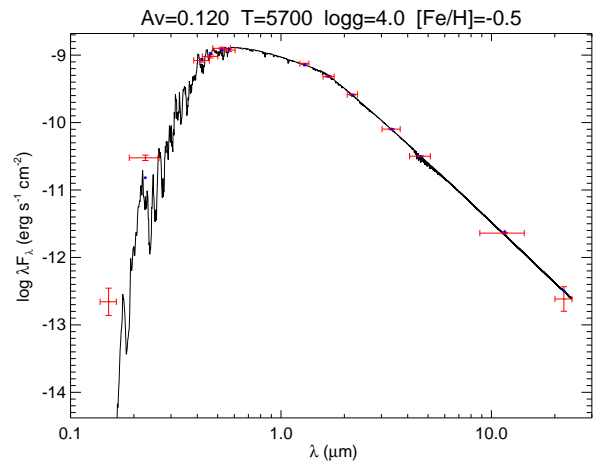}
\caption{\label{fig:sedfit}The observed SED from the near-UV through the IR for 
TYC~2087-00255-1, along with a best-fit NextGen model atmosphere. Blue 
points represent the expected fluxes in each band based on the model, red 
horizontal bars are the approximate bandpass widths, and red vertical bars 
are the flux uncertainties.  There is potentially some GALEX FUV excess 
indicating this star is chromospheric active. The resultant fundamental 
stellar parameters from this fit agree with the stellar parameters determined 
from the stellar spectra to within 1-$\sigma$. }
\end{figure}

The spectral energy distribution (SED) was constructed for TYC~2087-00255-1 
in Fig.~\ref{fig:sedfit} using near UV \citep[GALEX,][]{morrissey07}, optical 
\citep{hog00, kharchenko09}, near IR \citep[2MASS,][]{cutri03} and IR 
\citep[WISE,][]{wright10, cutri12} photometric data. These photometric data 
are presented in Table~\ref{tab:photometry}. 
The data were fit with fluxes from a NextGen model atmosphere \citep{hau99}. 
We limited the maximum line-of-sight extinction to be $\rm A_V < 0.15$ 
based on the analysis of dust maps by \citet{schlegel98}. 
The resultant parameters, $T_{\rm eff}=5700\pm200$ K, 
$\log (g)=4.0\pm0.5$, $\rm [Fe/H]=-0.5\pm0.5$ and $A_V=0.12_{-0.06}^{+0.03}$, 
agree within $1$-$\sigma$ of the results found via analysis of our high resolution spectroscopy.
In the above analysis, we did not constrain any of the fit parameters except for $\rm A_V$. We 
did another fit where we forced $T_{\rm eff}$, $\log (g)$ and $\rm [Fe/H]$ to the spectroscopically 
determined values, which provide a more robust estimate of $\rm{A_V} = 0.12\pm0.03$. Using this 
total extinction estimate, and adopting a V band bolometric correction $BC_{V}$ of $-0.19\pm0.02$ \citep{cox00},  
we estimate the distance to this system is $218\pm14$ pc.

\begin{table}[htbp]
\begin{center}
\caption{ Photometric Measurements of the Star TYC~2087-00255-1 \label{tab:photometry}}
\begin{tabular}{lcc}
\hline\hline
Parameter & Value & Note\\
\hline
$galFUV$ &  21.285$\pm$0.506 & GALEX (Morrissey et al. 2007) \\
$galNUV$ &   15.511$\pm$0.013 & GALEX (Morrissey et al. 2007)\\
$B$ &    11.203$\pm$0.059 & \citet{kharchenko09} \\
$BT$ &    11.32$\pm$  0.06 & \citet{hog00}\\
$V$   &  10.553 $\pm$   0.048& \citet{kharchenko09} \\
$VT$  &  10.58 $\pm$    0.04  & \citet{hog00} \\
$J2M$  & 9.29 $\pm$     0.02 & \citet{cutri03} \\
$H2M$  & 8.96 $\pm$     0.03 & \citet{cutri03}  \\
$K2M$  &  8.88   $\pm$   0.02  & \citet{cutri03}  \\
$WISE1$ & 8.839 $\pm$   0.024 & \citet{cutri12}  \\
$WISE2$ & 8.864 $\pm$    0.022 & \citet{cutri12} \\
$WISE3$ & 8.792 $\pm$    0.027  & \citet{cutri12}\\
$WISE4$ & 9.169 $\pm$    0.454 & \citet{cutri12} \\
\hline
\end{tabular}
\end{center}
\end{table}



%




\subsection{Stellar Mass and Radius}
\label{sec:mass}

We determine the mass and radius of the parent star, TYC~2087-00255-1, from $T_{\rm
eff}$, $\log(g)$, and [Fe/H] using the empirical polynomial relations
of \citet{Torres2010}, which were derived from a sample of eclipsing
binaries with precisely measured masses and radii.  We estimate the
uncertainties in $M_*$ and $R_*$ by propagating the uncertainties in $T_{\rm
eff}$, $\log(g)$, and [Fe/H] (see Table~\ref{tab:stellarparams}) using
the covariance matrices of the \citet{Torres2010} relations (kindly provided by
G. Torres). Since the polynomial relations of \citet{Torres2010} were derived empirically, 
the relations were subject to some intrinsic scatter, which we add in quadrature to the 
uncertainties propagated from the stellar parameter measurements 
\citep[$\sigma_{\log m} = 0.027$ and $\sigma_{\log r} = 0.014$;][]{Torres2010}. 
The final stellar mass and radius values obtained are 
$M_* = 1.16 \pm 0.11\,M_{\odot}$ and $R_* = 1.64\pm0.37\,R_{\odot}$ 
(see Table~\ref{tab:stellarparams}).

\subsection{Evolutionary State}
\label{sec:evolution}
\begin{figure}
\includegraphics[angle=90,scale=0.35]{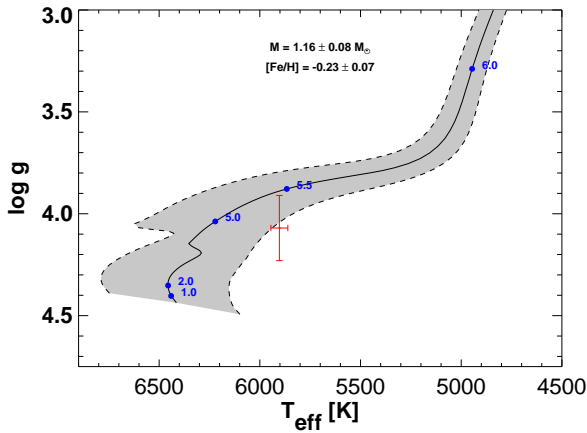}
\caption{\label{fig:evolution}The evolutionary track for an object with 
$M=1.16\,M_\odot$, at $\rm [Fe/H]=-0.23$.  Ages of 1, 2, 5, 5.5, 
and 6\,Gyr are indicated as dots. The possible tracks for $\pm 1\sigma$ 
deviation in the mass are shown by the shaded region.  The stellar parameters 
for TYC~2087-00255-1, with 1-$\sigma$ error bars, are shown by the cross.}
\end{figure}

In Fig.~\ref{fig:evolution} we compare the spectroscopically measured $T_{\rm eff}$ and $\log (g)$ 
of TYC~2087-00255-1 (red error bars) against a theoretical stellar evolutionary track from the 
Yonsei-Yale (``Y$^2$'') model grid \citep{demarque04}. The solid curve represents 
the evolution of a single star of mass $1.16 \pm 0.11\,M_{\odot}$ and metallicity of 
$\rm [Fe/H]=-0.23\pm0.07$. The dashed curves represent the same evolutionary track but 
for masses $\pm 0.08\,M_{\odot}$, which represents the $1$-$\sigma$ uncertainty in our 
derived mass. The filled gray region between the mass tracks represents the 
expected location of a star of TYC~2087-00255-1's mass and metallicity as it evolves off 
the main sequence.  The spectroscopically measured $T_{\rm eff}$, $\log(g)$, and 
$\rm [Fe/H]$ place TYC~2087-00255-1 in the subgiant phase, prior to the base of 
the red giant branch, with an estimated age of $\sim5.5$ Gyr.

\subsection{Stellar Rotation Period and Rotational Velocity}

\label{sec:obsphotometry}



\begin{figure}
\plotone{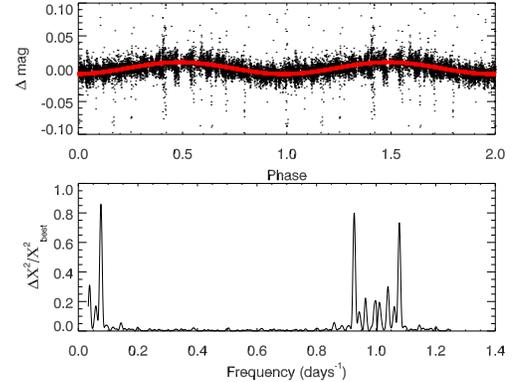}
\caption{\label{fig:wasp}
{\bf Top:}  Phase-folded light curve for TYC~2087-00255-1 at a period of 13.16 day 
from SuperWASP.  {\bf Bottom:}  
Lomb-Scargle periodogram of the SuperWASP data,
showing no evidence for any significant periodicities
around $P=9$\,days (frequency of 0.11 days$^{-1}$, the orbital period of MARVELS-4b). Instead there is 
evidence of a significant period at $13.16$\,days (frequency of 0.076 days$^{-1}$), which is likely the rotational 
period of TYC~2087-00255-1 as tracked by rotational modulation of starspots. There are several peaks around the $1$-day period, which are daily aliases.}
\end{figure}

In this section, we will use three different ways to estimate the rotation 
period of TYC~2087-00255-1. First, 
we find a sinusoidal variation in the 2004 SuperWASP photometry 
data with a period $P = 13.16 \pm0.01$ days and amplitude of 9 mmag. 
In Fig.~\ref{fig:wasp} we have shown a phase-folded plot 
of the 2004 SuperWASP data. Since the SuperWASP photometry 
data used an aperture radius of 49 arcseconds, we used 
the SIMBAD website to check for bright stars inside this 49 arcseconds 
aperture which could account for this 9 mmag variation and 
found several stars with $V_{mag} > 16.2$. Such faint stars 
could not produce such a 9 mmag variation around the 
$V_{mag}=10.6$ TYC~2087-00255-1 unless their luminosities 
varied by $100\%$, which is quite unlikely. We can explain 
this $13.16$~d period as the rotation period of the host star 
and the 9 mmag sinusoidal variation by the rotational 
modulation of star spots.

The second method to derive the rotation period is to use 
the chromospheric activity-rotation relation \citep{mamajek08}. 
The chromospheric Ca~II HK activity index of TYC~2087-00255-1 
derived previously is $\log R^{'}_{\rm HK}=-4.58$. The 
corresponding rotation period is $11.8$~d, 
with an estimated error of $~2.4$~d from this calibrated relation. 
This rotation period agrees with the one derived above ($13.16$~d) 
using the SuperWASP photometry data. However, we should 
note here that this chromospheric activity-rotation relation is 
derived for solar-type dwarf stars. Since TYC~2087-00255-1 
is a slightly evolved subgiant, it may not be appropriate to use 
this relation here.

Thirdly, we use the equation $2\pi R_*/v_{rot}\sin i$ to 
estimate the rotation period of the host star. 
We measured the projected rotational velocity $v_{rot}\sin i$ 
of TYC~2087-00255-1 by comparing our high resolution 
SARG and FEROS spectrum to broadened versions of 
Kurucz ATLAS synthetic spectra. We used the atmospheric parameters 
derived in \S 3.1 and fixed the macro turbulence velocity to 
$V_{macro} = 4.2$ km s$^{-1}$ based on Equation (1) from \citet{valenti05}. 
We find $v_{rot}\sin i = 9.2\pm2.0$ km s$^{-1}$ using the observed SARG 
spectrum and $v_{rot}\sin i = 10.1\pm0.9$ km s$^{-1}$ using the observed 
FEROS spectrum. Using the derived radius of TYC~2087-00255-1 
$R_*=1.64\pm0.37 R_\odot$, the corresponding rotation period 
are $9.1\pm2.9$ d and $8.3\pm 2.0$ d. These are about 1.4-sigma 
and 2.4-sigma from the 13.16 day period inferred from photometric 
data, suggesting that the star is likely close to edge on, as a smaller 
inclination would imply shorter periods and thus larger discrepancy 
with the photometric period. The slight tension even assuming 
$\sin i =1$ may arise from systematic errors in the estimate of $v_{rot}\sin i$, 
including an incorrect assumed value for $V_{macro}$.



\section{TYC~2087-00255-1: The Companion} 
\subsection{Keplerian Orbital Solution}
\label{sec:kpo}

\begin{figure}
\plotone{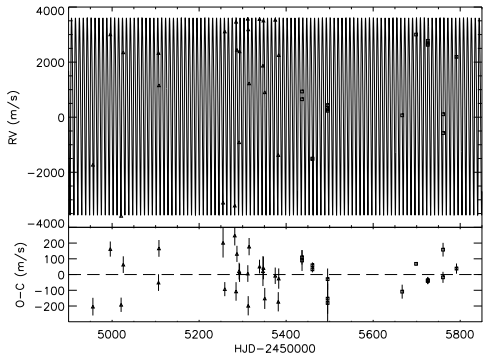}
\caption{\label{fig:rvcurve} {\bf Top:}  Keplerian orbital 
solution for TYC~2087-00255-1. Open triangles are MARVELS discovery 
data, open squares are TNG data. {\bf Bottom:} The residuals between 
the data points and the orbital solution.}
\end{figure}

\begin{figure}
\plotone{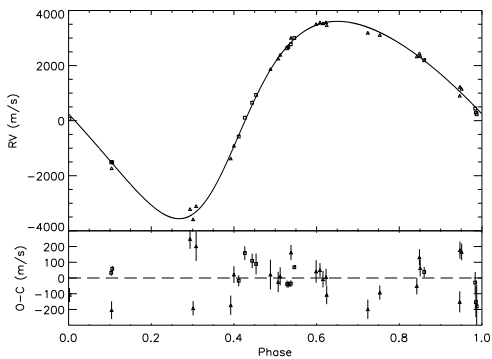}
\caption{ \label{fig:rvcurve_phased} 
{\bf Top:} Phase-folded Keplerian orbital solution and RV 
residuals for TYC~2087-00255-1.  Open triangles are MARVELS 
discovery data, open squares are TNG data.  {\bf Bottom:} The residuals 
between the data points and the orbital solution. }
\end{figure}

Radial velocities derived from MARVELS and SARG data were used to fit 
Keplerian orbital parameters. Since there are star spot features seen from the light curve 
of TYC~2087-00255-1 (see \S \ref{sec:obsphotometry}), we added an additional 
stellar `jitter' term ($\sigma_{\rm jitter}$) in our Keplerian orbital model as 
suggested by \citet{Ford06} to account for any additional noise induced by the stellar activity.  
We performed two Keplerian orbital fits using the Markov Chain Monte Carlo 
method \citep[MCMC, see, e.g.,][]{Ford06}. First, we use only the MARVELS 
data to fit the Keplerian orbit, then we combine the MARVELS and SARG RV data 
to do a joint fit. For the details of our one planet/BD 
RV model, please see \S 2 of \citet{gregory07}. The best-fit parameters 
from the two fits are presented in Table~\ref{tab:companionparams} and they 
agree with each other quite well. 


As expected, the fit of the combined MARVELS and SARG RV data has relatively 
smaller uncertainties compared to the fit with just the MARVELS data, therefore we use this 
combined fit as our final Keplerian orbital solution. In this fit, a constant systematic 
velocity term is included for each of the two instruments to account for the offset between the 
observed differential RV data and the zero-point of the Keplerian RV model
($-0.803\pm0.035$~$\rm{km~s^{-1}}$ for MARVELS and $-353\pm0.029$~$\rm{km~s^{-1}}$
for SARG). The RV data shown in Table~\ref{tab:obsjournalhet1} and Table~\ref{tab:obsjournalhet2} 
are the RVs after subtraction of these two constant systematic velocity terms.  
The final Keplerian orbit of MARVELS-4b has a period $P = 9.0090 \pm 0.0004$ days, 
 $e = 0.226 \pm 0.011$ and semi-amplitude $K = 3.571 \pm 0.041 \rm{km~s^{-1}}$. 
This solution is shown in Figs~\ref{fig:rvcurve} and \ref{fig:rvcurve_phased} together with the MARVELS and 
SARG RV data. The residuals shown in these two plots could not be explained 
only by the errors in our RV data. A stellar jitter term $\sigma_{\rm jitter}=112\pm17$m s$^{-1}$ 
is required in our fitting to explain these residuals. In our previous paper \citep{fleming10,lee11,wis12}, 
we did not include a stellar `jitter' term in our Keplerian fitting. Instead we 
re-scaled the error bar of MARVELS RV data to force the reduced 
chi-square to be 1. Since MARVELS-4b is orbiting an active star which 
has spots activity, we choose to include this `jitter' term and did not try to 
rescale the MARVELS and SARG RV error bar. We note here that both methods 
(including a `jitter' term or rescaling the RV error bar for the reduced chi-square to be 1)
yield Keplerian orbital parameters consistent with each other in the $1\sigma$ range. 
We will discuss more about this stellar jitter in \S \ref{sec:jitter}.
Using the derived value of 
$M_*$ in \S \ref{sec:mass}, we estimate a minimum mass (i.e., for $\sin \alpha=1$ 
where $\alpha$ is the line-of-sight orbital inclination) for the companion, MARVELS-4b, 
as $m_{\rm min}=40.0 \pm 2.5\,M_{Jup}$, where the uncertainty is 
dominated by the uncertainty in the primary mass. Assuming the system 
is edge on, the semimajor axis of this system is $a=0.090\pm0.003$\,AU.

\begin{table*}[htbp]
\begin{center}
\caption{{ MARVELS-4b:  Parameters of the Companion \label{tab:companionparams}}}
\begin{tabular}{ccc}
\hline\hline
Parameter & MARVELS+TNG & MARVELS \\
\hline
Minimum Mass & 40.0$\pm$ 2.5 $M_{Jup}$ & 39.9$\pm$ 2.5 $M_{Jup}$\\
$a$ & 0.090 $\pm$ 0.003 AU & 0.090 $\pm$ 0.003 AU \\
$K$ (km~s$^{-1}$)& 3.571 $\pm$ 0.041 km~s$^{-1}$ & 3.563$\pm$ 0.073 km~s$^{-1}$ \\
$P$ (d) & 9.0090 $\pm$ 0.0004 d & 9.0105 $\pm$ 0.0024 d  \\
$e$ & 0.226$\pm$  0.011 & 0.233$\pm$  0.022 \\
$w$ & 4.086$\pm$ 0.041 & 4.077$\pm$0.081\\
$\sigma_{\rm jitter}$ & 0.112$\pm$ 0.017 km~s$^{-1}$  & 0.152$\pm$ 0.046 km~s$^{-1}$ \\
$T_{\rm{prediction\; for\; transit}}$ (HJD$_{\rm UTC}$) & 2455549.629 $\pm$ 0.056 &  2455549.715 $\pm$ 0.082 \\
$T_{\rm periastron}$ (HJD$_{\rm UTC}$) &  2455552.797$\pm0.083$ & 2455552.851$\pm0.147$  \\
\hline
\end{tabular}
\end{center}
\end{table*}

\subsection{Search for Possible Transit Signals}
Based on the orbital parameters of
the companion, we have calculated the a priori transit
probability to be $7.8\%$ using Equation (5) from \citet{kane08}. 
The expected duration of a central transit is $\sim R_{*}P/(\pi a) = 2.94$ hr, 
and the expected depth is $\sim (R/R_*)^2 = 0.004(R/R_{\rm Jup})^2$, where $R$ is 
the radius of the companion. We phase folded the SuperWASP data and 
searched for a transit signal at the expected transit time, but no evidence for a 
transit with a period $\sim 9$ days was found. We could rule out existence of 
a transit signal with depth greater than 0.014 mag at 3-$\sigma$ confidence level. 
Our conclusion is that the SuperWASP data is not sensitive enough to detect a 
4 mmag transit signal. Further photometric follow-up observations are needed to 
rule out or confirm this transit signal.

\subsection{Search for Possible Stellar Companions Using Lucky Imaging}
\label{FC:DataAnalysis}
\begin{figure}
\plotone{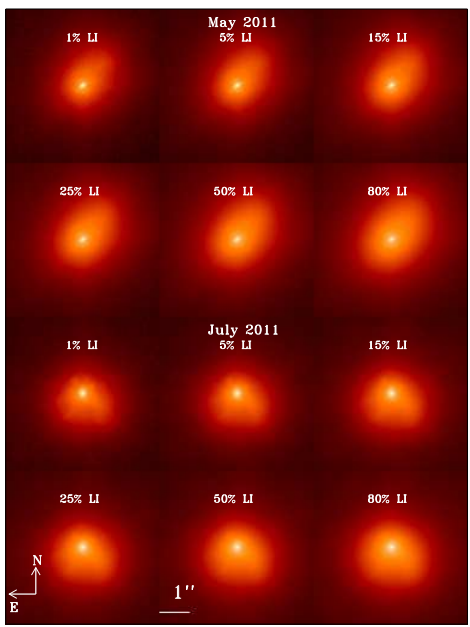}
\caption{Composite images showing the results of different LI thresholds applied to frames 
acquired with FastCam on May 8th 2011 and July 1st 2011.  
Each panel covers $\sim 5.5 \times 5.5$ square arcseconds, centered on TYC~2087-00255-1.
To ease visualization, each image has been normalized to a peak value of unity. This set of images illustrates the 
gain in angular resolution close to the target location when applying highly 
restrictive LI thresholds, but at the cost of lowering the contrast achieved 
at larger angular distances from target location. 
  \label{fig:LI_panels}}
\end{figure}

\begin{figure}
 \plotone{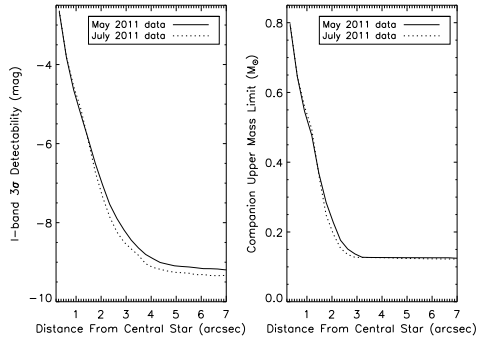}
 \caption{ {\it Left:} Best $3\sigma$ I-band detectability curves from lucky images achieved on 
 2011 July 1 and 2011 May 8. We see the quality of both nights are 
 comparable. 
 {\it Right:} Conversion of 3-$\sigma$ detectability curves into mass sensitivities 
 using empirical Mass-Luminosity relationships in the literature. 
 See the main text for details. \label{fig:MassConstraint}}
\end{figure}

We use the lucky images to search for, and place constraints on,  
any possible undetected stellar companions at large separations. 
The data were processed using a 
custom IDL software pipeline. After identifying corrupted frames due to cosmic 
rays, electronic glitches, etc., the remaining frames are bias corrected and flat fielded.
Lucky image selection is applied using a variety of selection thresholds (best $X$\%) 
based on the brightest pixel (BP) method. The selected BP must be below 
a specified brightness threshold to avoid selecting cosmic
rays or other non-speckle features.  As a further check, the BP must be 
consistent with the expected energy distribution from a diffraction 
speckle under the assumption of a diffraction-limited PSF.  
The frames are sorted from brightest to faintest according to the brightness of their BPs 
and the brightest $X$\% are then shifted and added to generate a final image. 
In the different panels of Fig.~\ref{fig:LI_panels}, we show the composite 
lucky images generated from considering only the best $\left\{1,5,15,25,50,80\right\}$\% of the image 
frames. Each panel covers $\sim 5.5 \times 5.5$ arcsec$^2$ centered on 
TYC~2087-00255-1. Restricting the images used to the best 1\% improves the angular resolution with 
respect to the loose LI selection (i.e. the top 80\% of the images image), but at the cost of increased 
noise at larger separations from the target. 
Following the same procedure as in \citet{fem2011}, we compute the 3-$\sigma$  ``best LI detectability curves'' 
for each of the two nights' data, which are shown in the left panel of Fig.~\ref{fig:MassConstraint}.
We use the following procedure to place a 3-$\sigma$ upper limit on the mass of a possible 
undetected stellar companion:

\begin{enumerate}
\item From the SED fitting of TYC~2087-00255-1 in \S \ref{sec:params} that provides an 
estimate of the extinction and distance, we estimate the absolute $I$ band magnitude 
of TYC~2087-00255-1 to be $M_I=2.74$. \item Knowing $M_I$ and the 3-$\sigma$ 
detectability curves (see the left panel of Fig.~\ref{fig:MassConstraint}) allows 
us to construct the $M_I$ versus $\rho$ (angular distance 
from TYC~2087-00255-1) curve for TYC~2087-00255-1. This curve provides 
the upper limit of absolute $I$ band magnitude for any undetected stellar 
companion at the 3-$\sigma$ level.
\item Although TYC~2087-00255-1 is identified as a subgiant, the fact that it is not very massive 
makes it plausible that any stellar companion with the same age and smaller mass will still be a 
main sequence object. These facts justify the use of the conversion from $I$ to $V$ band in \citet{mam2010}.
\item From the $M_V$ versus $\rho$ curves we employ the empirical Mass-Luminosity 
relationships (MLRs) from the literature \citep{hen1999,del2000,hen2004, xia2008,xia2010} to 
derive the 3-$\sigma$ upper mass limit for any undetected companion as a function 
of angular distance to TYC~2087-00255-1. 
\end{enumerate}

The results of applying the above procedure to our LI data are shown in 
the right panel of Fig.~\ref{fig:MassConstraint}.
 
\subsection{Search for Possible Stellar Companions Using AO Imaging}
\label{sec:AO}

\begin{figure}
\plotone{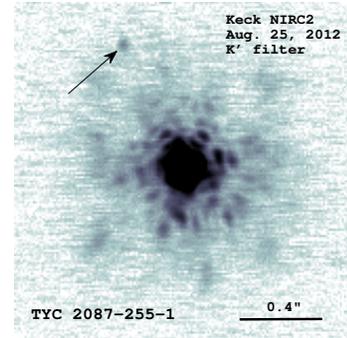}
\caption{Keck K' band AO image of TYC~2087-00255-1. A candidate 6.49 
mags fainter than the primary star in K’ band is identified. It has 
an angular separation of  $643\pm10$ mas and position 
angle $27.1\pm0.1$ degree (pointed by an arrow in the plot).  
  \label{fig:AO}}
\end{figure}

In this section we use the acquired Keck AO images to search for 
possible stellar companions around TYC~2087-00255-1. The AO images 
were processed by replacing hot pixel values, flat-fielding the array, 
and subtracting thermal background noise. No companions were identified in 
individual raw frames during the observations. However, upon stacking the 
images we noticed a point source to the north-east of TYC~2087-0025501. 
Fig.~\ref{fig:AO} shows the final processed K’ image. The candidate is 6.49 
mags fainter than the primary star in K’. We measure an angular separation 
of  $643\pm10$ mas and position angle $27.1\pm0.1$ degree. Assuming an 
age of 5 Gyr, the \citet{baraffe98} theoretical evolutionary models predict a mass 
of $0.13M_{\odot}$ if the candidate is physically associated at a distance 
of 218 pc. With a proper motion of $(\mu_\alpha\cos \delta, \mu_\delta) =(-2.9, 39.8)$ 
mas/yr for TYC~2087-00255-1\citep{hog00}, it will be possible to determine 
whether this candidate is a tertiary companion in less than one year with 
NIRC2.

%





\section{DISCUSSION}
\subsection{Does MARVELS-4b Reside in the Low-mass Tail of the Stellar Formation Process?}
The mass of BD overlaps both with low-mass stars formed  
from collapse/fragmentation of molecular cores and massive 
planetary companions formed in protoplanetary disks. It is still not clear 
which mechanism dominates the formation of BDs. By 
extrapolating the companion mass function from both the exoplanet side 
and low-mass star side, \citet{Grether06} find the minimum number of 
companions per unit interval in log mass is $31^{+25}_{-18}M_{\rm Jup}$. 
\citet{sahlmann11} see evidence for a bimodal distribution in BD 
masses, with the gap between 25 and 45 $M_{\rm Jup}$ almost entirely 
devoid of objects. They suggest that the less-massive group may 
represent the high-mass tail of the planetary distribution. If true, 
the maximum mass of giant ``planets'' should be around 
$30M_{\rm Jup}$. The BD candidate MARVELS-4b reported in this 
paper has a minimum mass of $40.0\pm 2.5 M_{\rm Jup}$, which 
suggests that MARVELS-4b more likely formed like stars through 
collapse and/or fragmentation of molecular cores.





\subsection{Activity, Rapid Rotation and Stellar Spin Evolution}
\label{sec:tidal}

TYC~2087-00255-1 is in its subgiant phase with an estimated age of
$\sim \! 5.5$\,Gyr and chromospheric Ca~II HK index $\log R^{'}_{\rm
HK}=-4.58$ (see \S \ref{sec:params}).
\citet{jenkins11} have studied chromospheric activity indices for 
more than 850 FGK-type dwarfs and subgiant stars, and find the distribution of activity indices 
($\log R^{'}_{\rm HK}$) for their subgiant sample can be fit by a Gaussian 
centered around $-5.14$, with $\sigma=0.06$. The activity index for TYC~2087-00255-1
is $\log R^{'}_{\rm HK}=-4.58$, which makes it an unusually active subgiant star (see 
also \S~\ref{sec:params}). 

For cool stars with $T_{\rm eff} < 6500$ K, chromospheric activity is 
generated through a stellar magnetic dynamo,
which is related to the rotational velocity and rotation period of the
star. The star's rotation period increases with age through mass loss
in a magnetized wind \citep[``magnetic braking"][]{schatzman62,weber67, mestel68, 
skumanich72, epstein12}, and as such, its chromospheric 
activity level is expected to also decay with age, a phenomenon that 
has been observed \citep{wilson63, skumanich72, soderblom91}. For an evolved star 
such as TYC~2087-00255-1, the rotation period should be relatively large, 
$\gsim 30$~days due to magnetic braking and conservation of 
angular momentum as the radius expands \citep{skumanich72, epstein12}. 
However, we find the rotation period ($13.16$~days) is much shorter than 
this value. One possible explanation is that the tidal interaction 
with the companion has spun up the star and keeps the stellar 
magnetic dynamo active. In this scenario, the tidal interaction 
transfers orbital angular momentum into stellar rotational angular 
momentum. We will explore the coupled roles of radial expansion and tidal 
evolution, and find that the observed state is consistent with 
tidal theory and stellar evolution.


For a star born with $T_{\rm eff} \lesssim 6500$ K, it could develop 
a convective zone and lose angular momentum through 
``magnetic braking". So even when it rotates quickly at birth, it 
will spin down quickly and thus have a longer period at ZAMS. 
While for a star born with $T_{\rm eff} \gtrsim 6500$ K, 
it will be fully radiative and could not lose angular momentum through 
``magnetic braking". Thus it will rotate much more quickly. 
From the evolution track in Fig.\ref{fig:evolution} we could see that 
the effective temperature of TYC~2087-00255-1 is initially around 
the transition point $\sim 6500$~K, and its evolution is
difficult to know {\it a priori}. If fully radiative, we expect it to
rotate quickly while on the main sequence, while if convective, 
it would have spun down and will rotate more slowly. 
Here we consider both scenarios and find that we cannot 
rule either out, at least using a simple, coupled
model for the expansion of the primary and the tidal evolution.

As star evolves, its radius expands and hence we expect the star to
spin down via conservation of angular momentum. If the rotational
frequency of the star is $\omega$, then its time rate of change due to
expansion is
\begin{equation}
\label{eq:spindown}
\frac{d\omega}{dt} \vert _{exp} = -\frac{2\omega}{R}\frac{dR}{dt}.
\end{equation}
The rate at which $\omega$ changes is therefore encapsulated in
$dR/dt$, which we can derive from stellar evolution models. Using the
``Y$^2$'' models \citep{demarque04}, we fit a third order polynomial to the radius as a
function of time:
\begin{equation}
\label{eq:R(t)}
\frac{R_*}{R_\odot} = 1.0275 + 0.1661t' - 0.0619t'^2 + 0.01147t'^3
\end{equation}
where $t'$ is the age of the star in Gyr. Differentiating with respect
to $t'$ we find
\begin{equation}
\label{eq:drdt}
\frac{dR_*}{dt'} = 0.1661 - 0.1238t' + 0.03441t'^2,
\end{equation}
from which we can solve Eq.~(\ref{eq:spindown}).

The tidal evolution is considerably more complicated as it depends on
many more parameters, as well as the tidal model employed. Here we use
the ``constant-phase-lag'' (CPL) model as described in
\cite{FerrazMello08}. This widely-used model assumes 
a constant phase offset between the location of the companion and
the tidal bulge. The magnitude of the phase lag is $\frac{1}{2Q}$, where $Q$
is the ``tidal quality factor.'' Different values of Q have been proposed to 
explain tidal evolution of different systems ($10^7$ in \citet{zahn89}; 
$10^5$ in \citet{meibom05};$10^6$ to $10^7$ in \citet{schlaufman10}; 
$10^8$ to $10^9$ in \citet{penev11}). 
The speed of the evolution is also a
function of the Love number of degree two, $k_2$, which is a measure of
the height of the tidal bulge. Rather than reproduce the set of six
coupled differential equations that comprises the CPL model, the reader
is referred to \cite{FerrazMello08}. Other tidal models exist
\cite[e.g.][]{Hut81,Leconte10,Hansen10} that make qualitatively
different assumptions. As we are only interested in demonstrating that
the observed configuration is consistent with tidal theory, we limit
our scope to the CPL model. We use the numerical methods
outlined in App.~E of \cite{Barnes12}.

We can use the best fit parameters in the CPL model, but we also must
set $Q$ and $k_2$, and the moment of inertia constant, or ``radius of
gyration'' $r_g$. A wide range of values have been proposed for $Q$,
with $10^6$ being a standard choice
\citep[\eg][]{Jackson08,Jackson09}. We set $k_2$ to 0.5, which is
arbitrary since the tidal evolution actually depends on the quotient of $Q$
and $k_2$, and currently we cannot disentangle the two. We set $R_g$
to 0.35, consistent with theoretical expectations for solar-like stars
\citep{ClaretGimenez90}. We assume the companion is tidally locked,
use its minimum mass, and has a radius of 1~$R_{Jup}$. We set both
bodies' obliquities to 0.  With these choices, we may integrate the
tidal evolution of the system forward, tracking the spins,
obliquities, orbital period, and orbital eccentricity.

Because the stellar rotational frequency depends on both its internal
and tidal evolution, we include both effects in our
model. Furthermore, we assume that both effects act independently, i.e. no
feedbacks are present. This sort of coupling has been applied to
stellar binaries before
\citep{ZahnBouchet89,KhaliulinKhaliulina11,Gomez12}, but this may be
the first time it has been done for a star with a BD companion.

\begin{figure}
\plotone{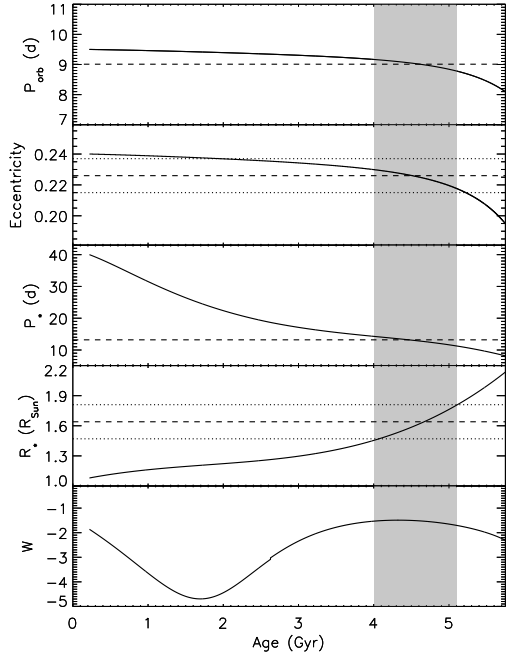}
\caption{Evolution of various properties of the MARVELS-4
system assuming a slow initial rotation period for the primary. {\it Top:} Orbital period. {\it Top Middle:}
Eccentricity. {\it Middle:} Primary's rotation period. {\it Bottom
  Middle:} Primary Radius. {\it Bottom:} Ratio of the time rate of
change of the primary's spin period from tides to that from expansion,
c.f. Eq.~(\ref{eq:spinratio}). \label{fig:slow} }
\end{figure}

We first consider the convective case, i.e. slow initial rotation.  In
Fig.~\ref{fig:slow}, we show one plausible history for this system. We
start the integration at an age of 221~Myr when the star is on the
main sequence, at which point the ``Y$^2$'' model \citep{demarque04}
predicts a radius of 1.08~$R_\odot$. The orbit begins with a period of
9.5 days, and an eccentricity of 0.24. The initial stellar rotation
period is 40 days. We set the stellar $Q$ to $4 \times 10^6$. Based on
the radial evolution and the observed uncertainties in radius, we
estimate the age of the system to be between 4 and 5.1 Gyr, and with a
nominal age of 4.75~Gyr, corresponding to the best fit radius of
$1.64~R_\odot$. This interval is shaded gray. This age estimate is
different from that in \S \ref{sec:evolution}, which is estimated from
$T_{\rm eff}$ and $\log (g)$ on the HR diagram. We also considered a
wider range of configurations and find that, for plausible primary spin
periods, the primary's $Q$ must lie in the range $3 \times 10^6$ -- $6
\times 10^6$ in order to produce a system consistent with the observations.

In the top panel, the evolution of the orbital period $P_{orb}$ is
shown by the solid curve. The dashed curve is the best fit orbit,
whose uncertainty is less than the curve thickness. The next panel
down shows the eccentricity evolution. The line styles are the same as
before, but now the $1\sigma$ observational uncertainty in $e$ is
denoted by the horizontal dotted lines. Next is the stellar rotation
period, shown in the same format as the previous panels. Fourth is the
evolution of the stellar radius, as given by the fit in
Eq.~(\ref{eq:R(t)}). Note that all these parameters pass through
their best fit values at our nominal age estimate.

The final panel shows the ratio of the stellar spin evolution from
tides to that from radial expansion (Eq.~[\ref{eq:spindown}]),
\begin{equation}
\label{eq:spinratio}
W \equiv \frac{ \frac{d\omega} {dt}\vert _{tides} }     { \frac{d\omega}{dt}\vert _{exp}  },
\end{equation}
where the denominator is the time rate of change of the rotational
frequency due to tides. The bottom panel of Fig.~\ref{fig:slow} shows the
evolution of $W$. Although $W$ evolves, it is always (1) negative, and
(2) more negative than $-1$. These features indicate that the tidal
torques oppose the radial expansion, and dominate. Therefore, the star
is spinning up, and will continue to do so until it becomes tidally
locked. (The discontinuity at $2.6~$Gyr is due to passage through the
$2:1$ spin-orbit resonance.)

\begin{figure}
\plotone{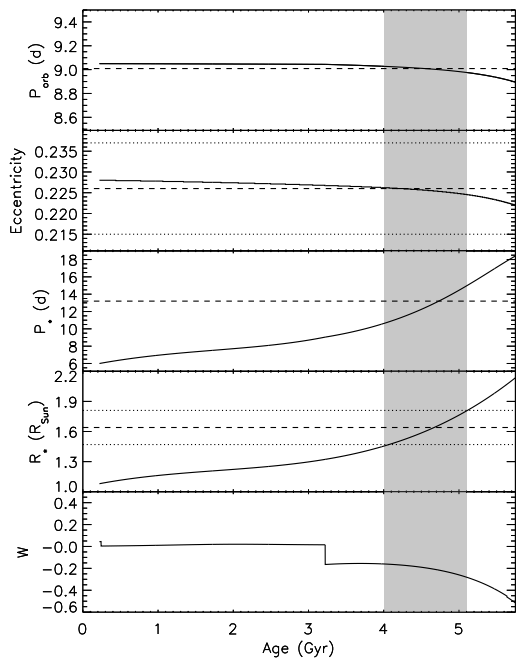}
\caption{Evolution of various properties of the MARVELS-4
system assuming a fast initial rotation period for the primary. The format is the same as Fig.~\ref{fig:slow}.\label{fig:fast} }
\end{figure}

We now turn to the possibility that the primary's rotation was
initially fast. At first glance, it may appear that the observed
system is inconsistent with such a history because the primary's
rotation period is currently longer than the orbital period, implying
the system has passed through the 1:1 spin-orbit resonance. In that
case, the primary may have become tidally locked. However, if the
expansion is rapid enough, and/or the tidal $Q$ large enough, the
primary could pass through this state and avoid permanent capture. In
Fig.~\ref{fig:fast}, we show such a configuration. In this case, the
primary's rotation period is initially 6 days, the eccentricity is
0.228, the primary's rotation period is 6 days, and its tidal $Q$ is
$3 \times 10^7$. As before the model predicts a system that could
evolve to the observed state.

The evolution of $W$ shows the complex evolution that may have
occurred. Initially $W > 0$ meaning that both tides and expansion act
in the same direction, increasing the rotational period in this
case. Very quickly the system passes through the 3:2 spin orbit
resonance and the phase lags change, producing the sudden drop in
$W$. For the next 2~Gyr $W$ increases slightly because the orbit is
shrinking and increasing the tidal torques. As the primary approaches
the 1:1 resonance, the torque decreases and $W$ decreases
accordingly. At 3.2~Gyr, the primary passes through the 1:1 resonance
and now the tides act to speed up the rotation and hence $W$ drops to
less than 0. Right after the resonance crossing, the tidal torques are
weaker, because the rotation is close to the equilibrium value (the
orbital period). However, as the primary continues to expand and slow
down, the disparity increases and the tidal torques grow larger. Thus,
$|W|$ increases because the tidal torque increases faster than radial
expansion slows the spin period. As $W$ never becomes more negative
than -1, the radial expansion dominates the evolution and rotational
slowdown continues.

We therefore have two competing, plausible evolutionary models for
this system. We have neglected some effects, such as magnetic breaking
and coupling between the expansion and tidal evolution, and hence we
do not express a preference for either model. However, given the
uniqueness of this system (evolving F star, BD/low mass stellar
companion, $W \sim 0$), it may be fertile ground for further
exploration and insight into tidal processes on stars with little or
no convective envelope.


In \S~\ref{sec:AO} we have found a point source near TYC~2087-00255-1 
using AO imaging. Interactions between MARVELS-4b and the 
tertiary may serve the purpose of bringing MARVELS-4b from its birth place 
to a tight orbit in the early history of this system through 
Kozai-Lidov mechanism if the tertiary is indeed associated with 
TYC~2087-00255-1 and if the initial mutual orbit inclination angle 
between MARVELS-4b and the tertiary is $39.2\degr\lesssim 
\delta_{23} \lesssim 141.8 \degr$ \citep{kozai62, lidov62}. 
This mechanism combined with tidal friction have been proposed 
to explain formation of close binaries in triple-star system \citep{mazeh79,kiseleva98, 
eggleton01, eggleton06, tokovinin06, fabrycky07} and formation of 
close-in Jupiter mass exoplanets \citep[``Hot Jupiters'';][]{wu03, fabrycky07,wu07,naoz11}. 
After MARVELS-4b has been brought to a tight orbit, tidal force 
from the primary takes over and MARVELS-4b follows the 
evolution illustrated above in Fig.~\ref{fig:slow} and Fig.~\ref{fig:fast} 
qualitatively. 


\subsection{Expected Stellar RV Jitter}
\label{sec:jitter}

Starspots and motions of the stellar surface are astrophysical sources of noise that 
can interfere with searches for companion RV signals.  These sources are commonly referred 
to as ``jitter'', first noticed by \citet{gunn79} and \citet{lupton87}, and subsequently explored 
by \citet{saar97}, \citet{saar98}, \citet{wright05}, \citet{lagrange09} and \citet{isaacson10}. 
Using the analytical relation given in \citet{saar97}, we estimate 
the expected RV jitter for TYC~2087-00255-1 as $6.5f^{0.9}v_{rot}\sin i$, 
where $f$ is the flux change in percent and $v_{rot}\sin i$ is the projected rotational velocity. 
We have derived $v_{rot}\sin i$ from our spectra using two different methods 
(see \S \ref{sec:obsphotometry}). The final combined value for $v_{rot}\sin i$ 
is $9.9\pm0.8$ km~s$^{-1}$ . The percentage change of the stellar flux 
is estimated to be $f \sim 1.8$ from the SuperWASP photometry data, 
thus the expected RV jitter is $\sim 106$ m~s$^{-1}$.


During the joint Keplerian orbital fit to the MARVEL+SARG RV data, 
a RV jitter $\sim 112$ m~s$^{-1}$ was needed to account for the extra noise 
in our RV measurements, which matches the expected jitter. 
However, when we use the MARVELS RV data only, 
we yield a `jitter' term $\sigma_{\rm jitter} =$ 152 m~s$^{-1}$ in our 
Keplerian orbital fit, which is bigger than the expected RV `jitter' arising from the 
stellar activity. This implies that uncharacterized systematics remain 
in the MARVELS RV data, and may likely dominate the stellar RV `jitter'.

\section{SUMMARY}

In a search through the first two years of SDSS-III MARVELS data, we
discovered MARVELS-4b, a candidate BD companion to the $V \! \simeq \! 10.6$
star TYC~2087-00255-1 with a velocity semi-amplitude of $K=3.571 \pm
0.041$\,km~s$^{-1}$ and a short orbital period of $9.0090 \pm
0.0004$\,d, yet with an eccentricity $e=0.226\pm0.011$.  
Additional RV data from SARG observations confirm the 
doppler variability. High-resolution spectroscopic observations 
indicate that the host star is a slightly evolved subgiant with $T_{\rm eff} = 5903\pm 42$\,K, 
$\log{g}= 4.07 \pm 0.16$, and [Fe/H]=$-0.23 \pm 0.04$, with an inferred
mass of $M_*= 1.16 \pm 0.11\,M_\odot$. The minimum mass of MARVELS-4b 
is $40.0 \pm 2.5\,M_{Jup}$, implying that it is most likely in the BD regime. 
A 13.16 day periodic signal is found in the SuperWASP photometry data, 
which is likely due to rotational modulation of starspots on the host star, 
and indicates that this star-BD system is not tidally synchronized. 
A second possible companion is found $643\pm10$ mas away 
from TYC~2087-00255-1 using K' AO imaging. Its association with the primary star 
could be verified by future proper motion measurements.
Ca~II H and K core emission indicates that the subgiant is chromospherically 
active at a level unusual for subgiants. Tidal interactions between the star and BD 
could have spun up the star and make it active. 

\acknowledgments

Funding for the MARVELS multi-object Doppler instrument was provided by the W.M. Keck 
Foundation and NSF with grant AST-0705139. The MARVELS survey was partially funded 
by the SDSS-III consortium, NSF grant AST-0705139, NASA with grant NNX07AP14G and 
the University of Florida. Funding for SDSS-III has been provided by the Alfred P. Sloan 
Foundation, the Participating Institutions, the National Science Foundation, and the U.S. 
Department of Energy Office of Science. The SDSS-III web site is \url{http://www.sdss3.org/}.
SDSS-III is managed by the Astrophysical Research Consortium for the Participating 
Institutions of the SDSS-III Collaboration including the University of Arizona, the Brazilian Participation Group, Brookhaven National Laboratory, University of Cambridge, Carnegie 
Mellon University, University of Florida, the French Participation Group, the German 
Participation Group, Harvard University, the Instituto de Astrofisica de Canarias, 
the Michigan State/Notre Dame/JINA Participation Group, Johns Hopkins University, 
Lawrence Berkeley National Laboratory, Max Planck Institute for Astrophysics, 
Max Planck Institute for Extraterrestrial Physics, New Mexico State University, 
New York University, Ohio State University, Pennsylvania State University, 
University of Portsmouth, Princeton University, the Spanish Participation Group, 
University of Tokyo, University of Utah, Vanderbilt University, University of Virginia, 
University of Washington, and Yale University.

Based on observations collected at Observat\'orio do Pico dos Dias (OPD), 
operated by the Laborat\'orio Nacional de Astrof\'{\i}sica, CNPq, Brazil.
FEROS spectra were observed at the ESO 2.2 m telescope under the ESO-ON agreement. 
This work has made use of observations taken with the Telescopio Nationale Galileo (TNG) 
operated on the island of La Palma by the Fundation
Galileo Galilei, funded by the Instituto Nazionale di Astrofisica (INAF), in
the Spanish {\it Observatorio del Roque de los Muchachos} of the Instituto de
Astrof{\'\i}sica de Canarias (IAC). 

This research is partially supported by funding from the Center for Exoplanets
and Habitable Worlds. The Center for Exoplanets and Habitable Worlds is supported by the
Pennsylvania State University, the Eberly College of Science, and the
Pennsylvania Space Grant Consortium. Keivan Stassun, Leslie Hebb, and Joshua 
Pepper acknowledge funding support from the Vanderbilt Initiative in Data-Intensive 
Astrophysics (VIDA) from Vanderbilt University, and from NSF Career award AST-0349075. 
EA thanks NSF for CAREER grant 0645416. GFPM acknowledges financial support from 
CNPq grant n$^{\circ}$ 476909/2006-6 and FAPERJ grant n$^{\circ}$ 
APQ1/26/170.687/2004. 
L.G. acknowledges financial support provided by the PAPDRJ CAPES/FAPERJ Fellowship.
J.P.W. acknowledges support from NSF Astronomy \& Astrophysics 
Postdoctoral Fellowship AST 08-02230.  LDF acknowledges financial support from CAPES.
RB acknowledges support from NSF AST grant 1108882. BSG 
acknowledges funding support from NSF CAREER grant AST-105652.

This research has benefitted from the M, L, and T dwarf compendium housed at DwarfArchives.org and maintained by Chris Gelino, Davy Kirkpatrick, and Adam Burgasser. 

We have used data from the WASP public archive in this research. The WASP consortium comprises of the University of Cambridge, Keele University, University of Leicester, The Open University, The Queen’s University Belfast, St. Andrews University and the Isaac Newton Group. Funding for WASP comes from the consortium universities and from the UK’s Science and Technology Facilities Council.

This publication makes use of data products from the Two Micron All Sky Survey, which is a joint project of the University of Massachusetts and the Infrared Processing and Analysis Center/California Institute of Technology, funded by the National Aeronautics and Space Administration and the National Science Foundation.

This publication makes use of data products from the Wide-field Infrared Survey Explorer, which is a joint project of the University of California, Los Angeles, and the Jet Propulsion Laboratory/California Institute of Technology, funded by the National Aeronautics and Space Administration.


\clearpage

\clearpage

\end{CJK}
\end{document}